\documentclass[aps,prd,twocolumn,superscriptaddress]{revtex4}
\usepackage{epsfig,epsf}
\usepackage{amsmath}
\usepackage{amsthm}
\usepackage{amsfonts}
\usepackage{amssymb}
\usepackage{dsfont}
\usepackage{multirow}
\usepackage{appendix}
\usepackage{slashed}
\usepackage[active]{srcltx}
\usepackage{psfrag}

\begin{document}

\title{The new $a_1(1420)$ state: structure, mass and width}
\date{\today}
\author{H.~Sundu}
\affiliation{Department of Physics, Kocaeli University, 41380 Izmit, Turkey}
\author{S.~S.~Agaev}
\affiliation{Institute for Physical Problems, Baku State University, Az--1148 Baku,
Azerbaijan}
\author{K.~Azizi}
\affiliation{Department of Physics, Do\v{g}u\c{s} University, Acibadem-Kadik\"{o}y, 34722
Istanbul, Turkey}
\affiliation{School of Physics, Institute for Research in Fundamental Sciences (IPM),
P.~O.~Box 19395-5531, Tehran, Iran}

\begin{abstract}
The structure, spectroscopic parameters and width of the resonance with
quantum numbers $J^{PC}=1^{++}$ discovered by the COMPASS Collaboration and
classified as the $a_1(1420)$ meson are examined in the context of QCD sum
rule method. In the calculations the axial-vector meson $a_1(1420)$ is
treated as a four-quark state with the diquark-antidiquark structure. The
mass and current coupling of $a_1(1420)$ are evaluated using QCD two-point
sum rule approach. Its observed decay mode $a_1(1420) \to f_0(980)\pi$, and
kinematically allowed ones, namely $a_1 \to K^{\ast \pm}K^{\mp}$, $a_1 \to
K^{\ast 0} \bar{K}^{0}$ and $a_1 \to \bar {K}^{\ast 0} K^{0}$ channels are
studied employing QCD sum rules on the light-cone. Our prediction for the
mass of the $a_1(1420)$ state $m_{a_{1}}=1416_{-79}^{+81}\ \mathrm{MeV}$ is
in excellent agreement with the experimental result. Width of this state $%
\Gamma=145.52 \pm 20.79 \mathrm{MeV}$ within theoretical and experimental
errors is also in accord with the COMPASS data.
\end{abstract}

\maketitle


\section{Introduction}

\label{sec:Intr }
Classification of the light scalar and axial-vector meson multiplets is
among of long-standing and ongoing problems of hadron spectroscopy \cite%
{Patrignani:2016xqp}. The conventional theory of mesons that considers them
as particles composed of a quark and antiquark $q\bar q$ meets with
abundance of observed light states that should be included into this scheme.
In fact, the family of axial-vector mesons contained till now five states
with the spin-parities $J^{PC}=1^{++}$ \cite{Chen:2015iqa}, recently
enlarged due to discovery by COMPASS Collaboration of a new resonance $%
a_1(1420)$ with the same quantum numbers \cite{Adolph:2015pws}. Some of
these states including $a_1(1420)$ meson were listed in Ref.\ \cite%
{Patrignani:2016xqp}, others wait detailed experimental explorations to
confirm their status. It seems new theoretical models are required to
explain all variety of available and forthcoming experimental information.

The COMPASS Collaboration analyzed the diffractive reaction $\pi^{-}+p \to
\pi^{-}\pi^{-}\pi^{+} +p_{\mathrm{recoil}} $ and studied $J^{PC}=1^{++}$
states in order to find a possible partner of the isosinglet $f_1(1420)$
meson. In the $f_0(980)\pi$ final state the Collaboration observed a
resonance $1^{++}$ and identified it as $a_1(1420)$ meson with the mass and
width
\begin{equation}
m=1414^{+15}_{-13} \ \mathrm{MeV},\ \Gamma=153^{+8}_{-23} \ \mathrm{MeV}.
\end{equation}

The discovery of the new light unflavored axial-vector state $a_{1}(1420)$
which presumably is isovector partner of $f_{1}(1420)$ meson, triggered
theoretical investigations in the context of different models aiming to
understand its quark-gluon structure and calculate its parameters . It is
interesting that by classifying $a_{1}(1420)$ as an axial-vector meson the
COMPASS Collaboration did not exclude its interpretation as an exotic state
\cite{Adolph:2015pws}. One of reasons pointed out there is observation of
only $a_{1}(1420)\rightarrow f_{0}(980)\pi $ decay mode of the new meson $%
a_{1}(1420)$. The abundance of axial-vector mesons in the mass interval $%
1.2\div 2\ \mathrm{GeV}_{{}}$, and difficulties in interpretation of $%
a_{1}(1420)$ as a radial excitation of the ground-state meson $a_{1}(1260)$
because of a small mass gap between them, also support attempts to interpret
it as an exotic state.

The final particle of the decay $a_1(1420) \to f_0(980)\pi$, namely the $%
f_0(980)$ meson provides an additional information on possible structure of
the $a_1(1420)$ meson. It is one of first mesons that was considered as
candidate for a light four-quark state. Thus, already at early years of the
quark-parton model it was supposed that the scalar meson $f_0(980)$ instead
of traditional $\bar q q $ structure may have $\bar{q}^2 q^2$ composition
\cite{Jaffe:1976ig}. Due to a substantial $s$-quark component it was treated
also as the $K\bar K$ molecule \cite{Weinstein:1990gu}. Lattice simulations
and experimental exploration seem confirm assumptions on the exotic nature
of $f_0(980)$ and some other hadrons \cite%
{Alford:2000mm,Amsler:2004ps,Bugg:2004xu,Klempt:2007cp}. Suggestions about a
diquark-antidiquark structure of the light scalar mesons including $f_0(980)$
one were made on the basis of new theoretical analysis in Refs.\ \cite%
{Maiani:2004uc,Hooft:2008we}, as well.

Sum rules studies of the light scalar nonet led to controversial conclusions
on their nature \cite%
{Latorre:1985uy,Narison:1986vw,Brito:2004tv,Wang:2005cn,
Chen:2007xr,Lee:2005hs,Sugiyama:2007sg,Kojo:2008hk,Wang:2015uha}. Thus,
calculations carried out in some of these works confirmed the
diquark-antidiquark structure of the light scalar particles \cite%
{Brito:2004tv,Wang:2005cn,Chen:2007xr}, whereas in Ref.\ \cite{Lee:2005hs}
an evidence for a diquark component in the light scalar mesons was not
found. Mixing of various diquark-antidiquarks with different flavor
structures \cite{Chen:2007xr}, superposition of diquark-antidiquark and $q%
\bar{q}$ constituents \cite{Sugiyama:2007sg,Kojo:2008hk,Wang:2015uha} were
examined to understand internal organization and explain experimental
features of the light scalars.

It is seen that different theoretical models consider the $f_{0}(980)$ meson
predominantly as a tetraquark state, or at least as a particle containing
substantial four-quark component. This circumstance alongside with above
arguments may give one a hint on possible exotic structure of the master
particle $a_{1}(1420)$ itself. Really, soon after observation of the $%
a_{1}(1420)$ meson various scenarios that treated it as an exotic state
appeared in literature. In Ref.\ \cite{Wang:2014bua} the $a_{1}(1420)$ meson
was realized as superposition of diquark-antidiquark and two-quark states.
The mass of this compound, in accordance with conclusions of Ref.\ \cite%
{Wang:2014bua} agrees with experimental data of the COMPASS Collaboration.
The $a_{1}(1420)$ meson as a pure diquark-antidiquark state was explored in
Ref.\ \cite{Chen:2015fwa}, results of which are in accord with data, as
well. It is worth noting that in both of these works QCD two-point sum rule
method were used.

Another confirmation of the multi-quark nature of $a_{1}(1420)$ came from
studies carried out in Ref.\ \cite{Gutsche:2017oro} within the soft-wall
AdS/QCD approach, where Schrodinger-type equation for the tetraquark wave
function was derived and solved analytically. The prediction for the mass of
the $J^{PC}=1^{++}$ tetraquark state obtained there agrees with data of
Ref.\ \cite{Adolph:2015pws}.

Alternative explanations of $a_{1}(1420)$ as manifestation of dynamical
rescattering effects in $a_{1}(1260)$ meson's decays are presented in the
literature by a number of papers \cite%
{Ketzer:2015tqa,Liu:2015taa,Aceti:2016yeb,Basdevant:2015wma,Wang:2015cis}.
In fact, the resonance in the $f_{0}(980)\pi $ final state was explained in
Ref.\ \cite{Ketzer:2015tqa} as a triangle singularity appearing in the
relevant decay mode of the $a_{1}(1260)$ meson. In accordance with the
proposed scheme this decay proceeds through three stages: at the fist step $%
a_{1}(1260)$ decays to a pair of $K^{\ast }\bar{K}$-mesons, at the second
stage $K^{\ast }$ meson decays to $K$ and $\pi $. At the final phase $K$ and
$\bar{K}$ combine to create the $f_{0}(980)$ meson. Analysis of these
processes and calculation of corresponding triangle diagram using the
effective Lagrangian approach reveals a singularity that may be interpreted
as the resonance seen by the COMPASS Collaboration. The same ideas were
shared by Ref.\ \cite{Liu:2015taa}, where manifestation of the anomalous
triangle singularity were analyzed in various processes, including $%
a_{1}(1260)\rightarrow f_{0}(980)\pi $ decay.

Recently, problems of the two-body strong decays of $a_1(1420)$ were
addressed in Ref.\ \cite{Gutsche:2017twh}. In this work partial decay
channels of $a_1(1420)$ were investigated in the framework of the covariant
confined quark model by treating $a_1(1420)$ as a tetraquark with both
diquark-antidiquark and molecular structures. When considering the decay $%
a_1(1420) \to f_0(980)\pi$ the final-state meson $f_0(980)$ was also chosen
as the tetraquark state with molecular or diquark composition. The partial
decay widths, and the full width of the $a_1(1420)$ state calculated in this
paper allowed authors to conclude that it is a four-quark state, and a
molecular configuration for $a_1(1420)$ is preferable than the
diquark-antidiquark structure.

As is seen, theoretical interpretations of the axial-vector state $%
a_{1}(1420)$ can be divided into two almost equal classes: in some of papers
it is treated as a four-quark system with different structures, in others -
considered as dynamical rescattering effect seen in the decay $%
a_{1}(1260)\rightarrow f_{0}(980)\pi $.

In the present work we are going to test $a_{1}(1420)$ as an axial-vector
diquark-antidiquark state, and at the first phase of our investigations,
calculate its mass and current coupling. To this end, we will make use of
QCD two-point sum rule approach by taking into account various vacuum
condensates up to dimension twelve \cite{Shifman1,Shifman2}. We will
afterwards investigate two-body strong decay channels of the $a_{1}(1420)$
meson and compute strong couplings $g_{[\ldots ]}$ corresponding to the
vertices $a_{1}f_{0}\pi $, $a_{1}K^{\ast \pm }K^{\mp },a_{1}K^{\ast 0}\bar{K}%
^{0}$ and $a_{1}\bar{K}^{\ast 0}K^{0}$. These couplings are crucial for
calculation of the partial width of the decay modes $a_{1}\rightarrow
f_{0}(980)\pi ,\ a_{1}\rightarrow K^{\ast \pm }K^{\mp },\ K^{\ast 0}\bar{K}%
^{0}$ and $\ \bar{K}^{\ast 0}K^{0}$.

The strong couplings will be computed in the framework of the QCD light-cone
sum rule (LCSR) approach \cite{Balitsky:1989ry,Belyaev:1994zk}, which is one
of the most reliable and universal nonperturbative methods to explore
hadrons' spectroscopic parameters and their decay modes. We will study the
decay $a_{1}\rightarrow f_{0}(980)\pi $ by treating the $f_{0}(980)$ meson
as the diquark-antidiquark state and evaluate the strong coupling $%
g_{a_{1}f_{0}\pi }$ in the context of the full LCSR method. The reason is
that after contracting quark fields from interpolating currents for $%
a_{1}(1420)$ and $f_{0}(980)$ a relevant correlation function depends on
distribution amplitudes (DAs) of the pion. In the case of vertices composed
of the $a_{1}(1420)$ state and two conventional mesons one has to apply LCSR
method in conjunction with technical tools of the soft-meson approximation.
In the soft approximation a correlation function instead of DAs depends on
local matrix elements of a final light meson, and as a result, to preserve
four-momentum conservation at a vertex one should set its momentum equal to
zero \cite{Agaev:2016dev}. Additionally, to remove unsuppressed terms from
the phenomenological side of sum rules one has to perform operations
explained in Refs.\ \cite{Belyaev:1994zk,Ioffe:1983ju}. The LCSR method and
soft approximation were adapted to study vertices involving a tetraquark and
conventional mesons in our work \cite{Agaev:2016dev}, and was later applied
to investigate decays of various tetraquarks \cite%
{Agaev:2016dsg,Agaev:2017uky,Agaev:2017foq,Agaev:2017tzv,Agaev:2017oay}.

The present work is structured in the following manner: In Sec.\ \ref%
{sec:Mass} we calculate the mass and current coupling of the $a_{1}(1420)$
meson by treating it as a diquark-antidiquark state. These parameters will
be used later to calculate widths of the $a_{1}(1420)$ meson's strong decay
channels. In the next Section we examine the vertex $a_{1}(1420)f_{0}(980)%
\pi \equiv a_{1}f_{0}\pi $, and calculate the strong coupling $%
g_{a_{1}f_{0}\pi }$ and width of the $P$-wave decay $a_{1}(1420)\rightarrow
f_{0}(980)\pi $. Section \ref{sec:DecK} is devoted to exploration of the $S$%
-wave processes $a_{1}\rightarrow K^{\ast \pm }K^{\mp },\ K^{\ast 0}\bar{K}%
^{0}$ and $\ \bar{K}^{\ast 0}K^{0}$, where we present sum rule predictions
for relevant strong couplings and partial decay widths. Here we also
determine full width of the $a_{1}(1420)$ state. Section \ref{sec:Notes}
contains our concluding notes. The light quark propagators and lengthy
analytical expression for the correlation function $\Pi _{\mathrm{V}%
}(M^{2},s_{0})$ are written down in Appendix.


\section{Mass and current coupling of $a_1(1420)$}

\label{sec:Mass}
The $a_1(1420)$ state is a neutral isovector meson with $%
I^{G}J^{PC}=1^{-}1^{++}$. In the diquark picture its quark content has the
form $([us][\bar u \bar s]-[ds][\bar d \bar s])/\sqrt 2$, whereas the
isoscalar partner of $a_1(1420)$, namely $f_1(1420)$ will have the
composition $([us][\bar u \bar s]+[ds][\bar d \bar s])/\sqrt 2$. In the
chiral limit adopted in present paper the particles $a_1(1420)$ and $%
f_1(1420)$ have equal masses (see, Ref.\ \cite{Chen:2015fwa}).

The next problem is connected with the flavor structure of the interpolating
current. It may have symmetric, antisymmetric or mixed-symmetric type flavor
structures. In the present work for the $a_1(1420)$ meson we choose an
interpolating current that belongs to a class of currents with the
mixed-type flavor symmetry (detailed discussion of these questions was
presented in Ref.\ \cite{Chen:2015fwa})
\begin{equation}
J_{\mu}(x)=\frac{1}{\sqrt{2}} [J_{\mu}^{u}(x)-J_{\mu}^{d}(x)] ,
\label{eq:Curr1}
\end{equation}%
where
\begin{eqnarray}
&&J_{\mu}^{q}(x)= q_{a}^{T}(x)C\gamma _{5}s_{b}(x) \left[ \overline{q}%
_{a}(x)\gamma _{\mu }C\overline{s}_{b}^{T}(x)\right.  \notag \\
&&\left. -\overline{q}_{b}(x)\gamma _{\mu }C\overline{s}_{a}^{T}(x)\right
]+q_{a}^{T}(x)C\gamma _{\mu}s_{b}(x)  \notag \\
&&\times\left[ \overline{q}_{a}(x)\gamma _{5} C\overline{s}_{b}^{T}(x) -%
\overline{q}_{b}(x)\gamma _{5}C\overline{s}_{a}^{T}(x)\right ].
\label{eq:Curr2}
\end{eqnarray}%
In Eq.\ (\ref{eq:Curr2}) $q$ is one of the light $u,\ d$ quarks, $a,b$ are
color indices and $C$ is the charge conjugation operator.

Having fixed the current $J_{\mu }(x)$ we start to calculate the correlation
function which enables us to extract the mass $m_{a_{1}}$ and coupling $%
f_{a_{1}}$ of the $a_{1}(1420)$ state. The correlation function necessary
for our purposes is given by the expression:
\begin{equation}
\Pi _{\mu \nu }(p)=i\int d^{4}xe^{ip\cdot x}\langle 0|\mathcal{T}\{J_{\mu
}(x)J_{\nu }^{\dagger }(0)\}|0\rangle .  \label{eq:CorrF1}
\end{equation}

In accordance with usual prescriptions of QCD sum rules one has to express $%
\Pi _{\mu \nu }(p)$ in terms of physical parameters of the axial-vector
particle(s). We use here \textquotedblleft a ground-state + continuum"
approximation by supposing that the current $J_{\mu }(x)$ couples to only $%
a_{1}(1420)$ meson. In this framework $\Pi _{\mu \nu }^{\mathrm{Phys}}(p)$
takes the following form

\begin{equation}
\Pi _{\mu \nu }^{\mathrm{Phys}}(p)=\frac{m_{a_{1}}^{2}f_{a_{1}}^{2}}{%
m_{a_{1}}^{2}-p^{2}}\left( -g_{\mu \nu }+\frac{p_{\mu }p_{\nu }}{p^{2}}%
\right) +\ldots   \label{eq:CorM}
\end{equation}%
where the dots indicate contributions of the higher resonances and continuum
states. To derive $\Pi _{\mu \nu }^{\mathrm{Phys}}(p)$ we use the  matrix
element of the  $a_{1}(1420)$ meson
\begin{equation}
\langle 0|J_{\mu }|a_{1}(p)\rangle =f_{a_{1}}m_{a_{1}}\epsilon _{\mu },
\label{eq:MEl1}
\end{equation}%
where $m_{a_{1}},$  $f_{a_{1}}$and $\epsilon _{\mu }$ are  the mass, the
coupling and  the polarization vector of the $a_{1}(1420)$ state,
respectively.

The next phase of studies implies calculation of the correlation function $%
\Pi _{\mu \nu }(p)$ using the quark-gluon degrees of freedom. This means
that one has to substitute the interpolating current defined by Eqs.\ (\ref%
{eq:Curr1}) and (\ref{eq:Curr2}) into Eq.\ (\ref{eq:CorrF1}) and contract
quarks fields that generate light quark propagators, explicit expression of
which is moved to Appendix. The general form of the correlation function
then is:
\begin{equation}
\Pi _{\mu \nu }^{\mathrm{OPE}}(p)=\Pi _{\mathrm{V}}(p^{2})\left( -g_{\mu \nu
}+\frac{p_{\mu }p_{\nu }}{p^{2}}\right) +\Pi _{\mathrm{S}}(p^{2})\frac{%
p_{\mu }p_{\nu }}{p^{2}},  \label{eq:CorrF2}
\end{equation}%
where $\Pi _{\mathrm{V}}(p^{2})$ and $\Pi _{\mathrm{S}}(p^{2})$ are
invariant amplitudes receiving a contribution only from an axial-vector and
scalar particles, respectively. Because we are interested in parameters of
the axial-vector state $a_{1}(1420)$ we choose the Lorentz structure $g_{\mu
\nu }$ and corresponding function $\Pi _{\mathrm{V}}(p^{2})$ to derive
desired sum rules. To this end, we apply the Borel transformation to $\Pi
_{\mu \nu }^{\mathrm{Phys}}(p)$ and $\Pi _{\mu \nu }^{\mathrm{OPE}}(p)$,
pick out $\mathcal{B}\Pi _{\mathrm{V}}(p^{2})=\Pi _{\mathrm{V}}(M^{2})$ and
equate it to relevant piece from the phenomenological side of the sum rule
\begin{equation}
m_{a_{1}}^{2}f_{a_{1}}^{2}e^{-m_{a_{1}}^{2}/M^{2}}+\ldots =\Pi _{\mathrm{V}%
}(M^{2}).  \label{eq:SR1}
\end{equation}%
In terms of the spectral density $\rho (s)$ the Borel transformation of $\Pi
_{\mathrm{V}}(p^{2})$ takes  very simple form and reads
\begin{equation}
\Pi _{\mathrm{V}}(M^{2})=\int_{4m_{s}^{2}}^{\infty }ds\rho (s)e^{-s/M^{2}}.
\label{eq:BorelSpectForm}
\end{equation}%
In order to derive sum rules one has to subtract contributions of the higher
resonances and continuum states. This can be achieved by invoking of
assumption about the quark-hadron duality and replacing
\begin{equation}
\int_{4m_{s}^{2}}^{\infty }ds\rho (s)e^{-s/M^{2}}\rightarrow
\int_{4m_{s}^{2}}^{s_{0}}ds\rho (s)e^{-s/M^{2}},  \label{eq:SubrtF}
\end{equation}%
where $s_{0}$ is the continuum threshold parameter: It separates
ground-state and continuum contributions from each other. Alternatively,
Borel transform of $\Pi _{\mathrm{V}}(p^{2})$ may be calculated directly
from its expression avoiding intermediate steps. Then continuum subtraction
is fulfilled in terms that are $\sim (M^{2})^{n},\ n=1,\ 2,\dots $ \cite%
{Belyaev:1994zk}.

In the present paper we calculate $\Pi _{\mathrm{V}}(M^{2})$ by taking into
account various condensates up to dimension twelve. Contributions of terms
up to dimension eight are obtained utilizing relevant spectral densities,
remaining terms are found by means of direct Borel transformation. Sum rules
for the mass and coupling of the $a_{1}(1420)$ state can be obtained from
subtracted version of Eq.\ (\ref{eq:SR1}) by means of standard operations.
After continuum subtraction $\Pi _{\mathrm{V}}(M^{2},\ s_{0})$ acquires a
dependence also on the continuum threshold parameter $s_{0}$. The final
result for $\Pi _{\mathrm{V}}(M^{2},s_{0})$ is written down in Appendix.

The quark propagators, and as a result sum rules for the mass and coupling
depend on numerous parameters, values of which should be specified to
perform numerical computations. Below we list the mass of the $s$-quark, and
vacuum expectation values of the quark, gluon and mixed local operators
\begin{eqnarray}
&&m_{s}=128_{-6}^{+10}~\mathrm{MeV}  \notag \\
&&\langle \bar{q}q\rangle =-(0.24\pm 0.01)^{3}\ \mathrm{GeV}^{3},\ \langle
\bar{s}s\rangle =0.8\ \langle \bar{q}q\rangle ,  \notag \\
&&m_{0}^{2}=(0.8\pm 0.1)\ \mathrm{GeV}^{2},\ \langle \overline{q}g_{s}\sigma
Gq\rangle =m_{0}^{2}\langle \overline{q}q\rangle ,  \notag \\
&&\langle \overline{s}g_{s}\sigma Gs\rangle =m_{0}^{2}\langle \bar{s}%
s\rangle ,  \notag \\
&&\langle \frac{\alpha _{s}G^{2}}{\pi }\rangle =(0.012\pm 0.004)\,\mathrm{GeV%
}^{4},  \notag \\
&&\langle g_{s}^{3}G^{3}\rangle =(0.57\pm 0.29)\ \mathrm{GeV}^{6}.
\label{eq:Param}
\end{eqnarray}%
It is useful to note that for $m_{s}$ we use its value rescaled to $\mu =1\
\mathrm{GeV}$ \cite{Patrignani:2016xqp}.

Sum rules depend also on auxiliary parameters $M^{2}$ and $s_{0}$ the choice
of which has to satisfy standard restrictions. Performed analyses allow us
to fix the working regions for $M^{2}$ and $s_{0}$:
\begin{equation}
M^{2}\in \lbrack 1.4,\ 1.8]\ \mathrm{GeV}^{2},\ s_{0}\in \lbrack 2.4,\ 3.1]\
\mathrm{GeV}^{2}.  \label{eq:Wind}
\end{equation}

In Figs.\ \ref{fig:Mass} and \ref{fig:Coupl} we depict the sum rules results
for the mass and current coupling of the $a_{1}(1420)$ state as functions of
the Borel and continuum threshold parameters. As is seen, prediction for the
mass is rather stable against varying of both $M^{2}$ and $s_{0}.$ The
dependence of $f_{a_{1}}$ on the Borel parameter at fixed $s_{0}$ is very
weak, whereas its variations with $s_{0}$ are noticeable and generate
substantial part of theoretical errors.

For $m_{a_{1}}$ and $f_{a_{1}}$ we find:
\begin{equation}
m_{a_{1}}=1416_{-79}^{+81}\ \ \mathrm{MeV},\
f_{a_{1}}=(1.68_{-0.26}^{+0.25}\ )\cdot 10^{-3}\ \ \mathrm{GeV}^{4}.
\label{eq:Res1}
\end{equation}%
Our result for the mass of the $a_{1}(1420)$ state is in excellent agreement
with data of the COMPASS Collaboration. It agrees also with the mass of the $%
a_{1}(1420)$ meson obtained previously in the diquark-antidiquark picture in
Ref.\ \cite{Chen:2015fwa}
\begin{equation}
m_{a_{1}}=(1440\pm 80)\ \ \mathrm{MeV},\ f_{a_{1}}=(1.32\pm 0.35\ )\cdot
10^{-3}\ \ \mathrm{GeV}^{4}.  \label{eq:Res2}
\end{equation}%
Within the theoretical errors the coupling $f_{a_{1}}$ is compatible with
the result of Ref.\ \cite{Chen:2015fwa} : there is a large overlap region
between two predictions. It is worth noting that we have rescaled the
coupling $\widetilde{f}_{a_{1}}\equiv f_{a_{1}}m_{a_{1}}=$ $(1.9\pm 0.5\
)\cdot 10^{-3}\ \ \mathrm{GeV}^{5}$ in accordance with our definition Eq.\ (%
\ref{eq:MEl1}) using for this purpose $m_{a_{1}}$ from Eq.\ (\ref{eq:Res2}).
Numerical difference between two sets of parameters stems from some subleading
terms in the corresponding spectral densities, which however do not affect
considerably the final results, as is seen from Eqs.\ (\ref{eq:Res1}) and (\ref{eq:Res2}).

\begin{widetext}

\begin{figure}[h!]
\begin{center} \includegraphics[%
totalheight=6cm,width=8cm]{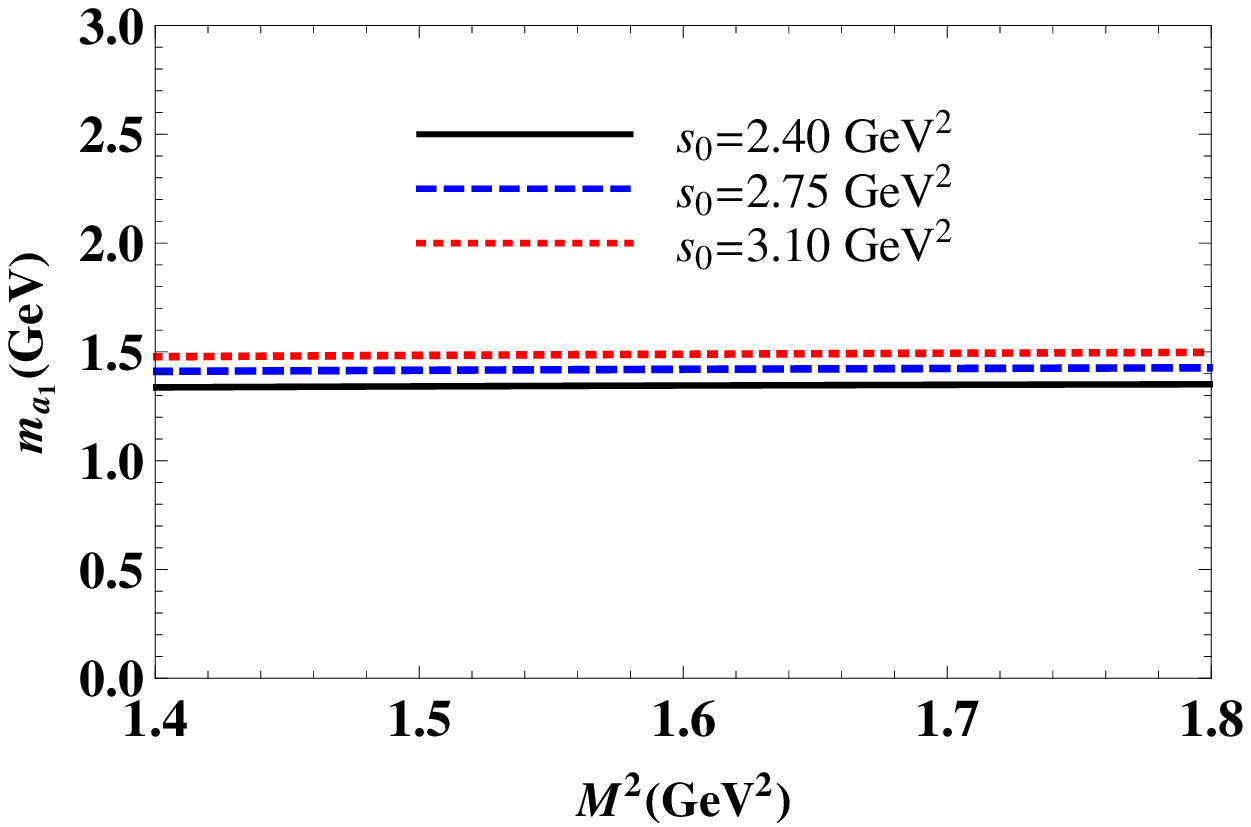}\,\,
\includegraphics[%
totalheight=6cm,width=8cm]{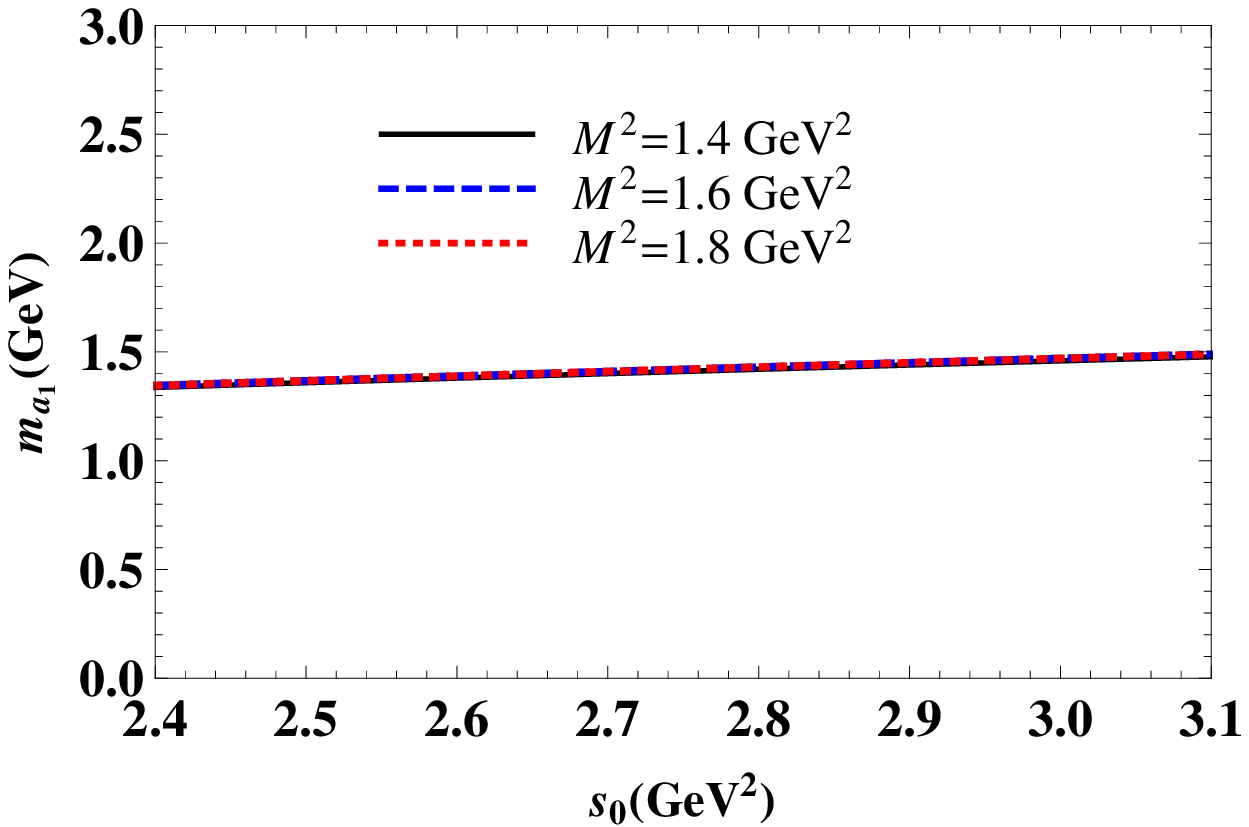}
\end{center}
\caption{ The mass
of the $a_1$ state as a function of the Borel parameter $M^2$ at fixed $s_0$
(left panel), and as a function of the continuum threshold $s_0$ at fixed $%
M^2$ (right panel).}
\label{fig:Mass}
\end{figure}
\begin{figure}[h!]
\begin{%
center}
\includegraphics[totalheight=6cm,width=8cm]{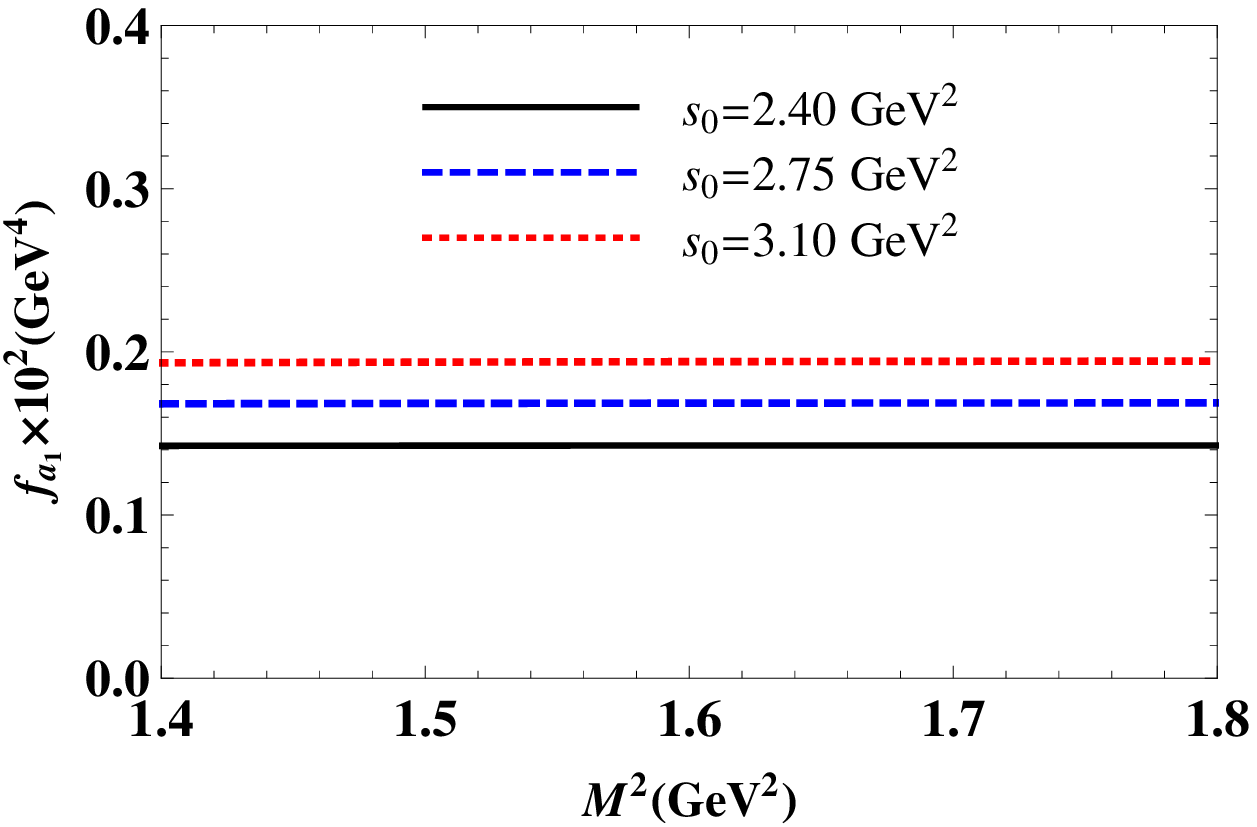}\,\,
\includegraphics[totalheight=6cm,width=8cm]{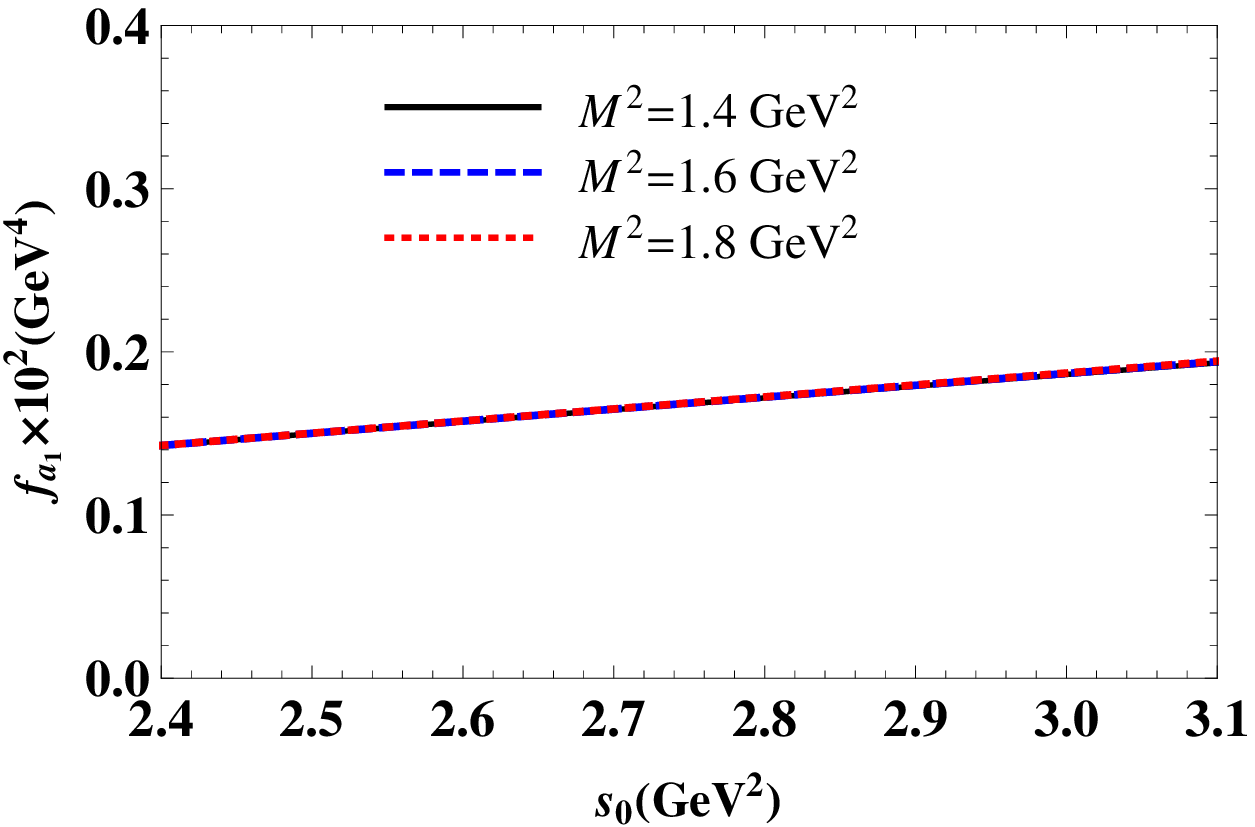} \end{center}
\caption{ The current coupling $f_{a_1}$ of the $a_1$ state as a function of $%
M^2$ at fixed $s_0$ (left panel), and of $s_0$ at fixed $M^2$ (right panel).}
\label{fig:Coupl}
\end{figure}

\end{widetext}

\section{The decay channel $a_1(1420) \to f_0(980) \protect\pi^0$}

\label{sec:Decf0}
The decay $a_{1}(1420)\rightarrow f_{0}(980)\pi ^{0}$ which was observed by
the COMPASS Collaboration and led to discovery of the axial-vector state $%
a_{1}(1420)$ is a $P$-wave transition and consequently is not its dominant
decay channel. Nevertheless, it should be properly analyzed because still
remains solely observed decay of the $a_{1}(1420)$ state . In the context of
the QCD light-cone sum rule method this process can be investigated starting
from the correlation function
\begin{equation}
\Pi _{\mu }(p,q)=i\int d^{4}xe^{ip\cdot x}\langle \pi (q)|\mathcal{T}%
\{J^{f}(x)J_{\mu }^{\dagger }(0)\}|0\rangle ,  \label{eq:CorrF3}
\end{equation}%
where $J_{\mu }(0)$ is the interpolating current of $a_{1}(1420)$ which we
treat in this work as the axial-vector diquark-antidiquark state, and $%
J^{f}(x)$ is the interpolating current of the scalar $f_{0}(980)$. In the
light of above discussions it is clear that there are various models of $%
f_{0}(980)$ in the literature. In the present study we consider $f_{0}(980)$
as the scalar diquark-antidiquark state and choose the current $J^{f}(x)$ to
interpolate it in the following form
\begin{eqnarray}
&&J^{f}(x)=\frac{\epsilon ^{dab}\epsilon ^{dce}}{\sqrt{2}}\left\{ \left[
u_{a}^{T}(x)C\gamma _{5}s_{b}(x)\right] \left[ \overline{u}_{c}(x)\gamma
_{5}C\overline{s}_{e}^{T}(x)\right] \right.  \notag \\
&&\left. +\left[ d_{a}^{T}(x)C\gamma _{5}s_{b}(x)\right] \left[ \overline{d}%
_{c}(x)\gamma _{5}C\overline{s}_{e}^{T}(x)\right] \right\} .
\label{eq:Curr3}
\end{eqnarray}%
After adopting the currents we analyze the vertex $a_{1}f_{0}\pi $ which is
composed of two tetraquarks and a conventional meson, and in this aspect
differs from ones containing a tetraquark and two ordinary mesons. In order
to derive the sum rule for the strong coupling $g_{a_{1}f_{0}\pi }$ we carry
out well-known standard operations. At the first stage we express the
correlation function in terms of physical parameters of the involved
particles and get
\begin{eqnarray}
&&\Pi _{\mu }^{\mathrm{Phys}}(p,q)=\frac{\langle 0|J^{f}|f_{0}(p)\rangle }{%
p^{2}-m_{f_{0}}^{2}}\langle f_{0}\left( p\right) \pi (q)|a_{1}(p^{\prime
})\rangle  \notag \\
&&\times \frac{\langle a_{1}(p^{\prime })|J_{\mu }^{\dagger }|0\rangle }{%
p^{\prime 2}-m_{a_{1}}^{2}}+...,  \label{eq:PhysDec1}
\end{eqnarray}%
where the dots indicate contributions due to higher resonances and continuum
states. The physical representation $\Pi _{\mu }^{\mathrm{Phys}}(p,q)$ of
the correlator can be simplified by means of the matrix element given by
Eq.\ (\ref{eq:MEl1}), a new one defined as
\begin{equation}
\langle 0|J^{f}|f_{0}\left( p\right) \rangle =f_{f_{0}}m_{f_{0}},
\label{eq:MEL2}
\end{equation}%
as well as by introducing the strong coupling $g_{a_{1}f_{0}\pi }$ to
specify the vertex
\begin{equation}
\langle f_{0}\left( p\right) \pi (q)|a_{1}(p^{\prime })\rangle
=g_{a_{1}f_{0}\pi }p\cdot \varepsilon ^{\prime \ast }.  \label{eq:Vert1}
\end{equation}%
Here $p^{\prime },\ p$ and $q$ are four-momenta of $a_{1}(1420),~f_{0}(980)$
and $\pi $, respectively. In Eq.\ (\ref{eq:Vert1}) $\varepsilon ^{\prime }$
is the polarization vector of the $a_{1}(1420)$ state. The two-variable
Borel transformations applied to $\Pi _{\mu }^{\mathrm{Phys}}(p,q)$ yield
\begin{eqnarray}
&&\mathcal{B}\Pi _{\mu }^{\mathrm{Phys}}(p,q)=g_{a_{1}f_{0}\pi
}m_{f_{0}}m_{a_{1}}f_{f_{0}}f_{a_{1}}e^{-m_{f_{0}}^{2}/M_{1}^{2}-m_{a_{1}}^{2}/M_{2}^{2}}
\notag \\
&&\times \left[ \frac{1}{2}\left( -1+\frac{m_{f_{0}}^{2}}{m_{a_{1}}^{2}}%
\right) p_{\mu }+\frac{1}{2}\left( 1+\frac{m_{f_{0}}^{2}}{m_{a_{1}}^{2}}%
\right) q_{\mu }\right] ,
\end{eqnarray}%
where $M_{1}^{2}$ and $M_{2}^{2}$ are the Borel parameters which correspond
to $p^{2}$ and $p^{\prime 2}$, respectively. The $\Pi _{\mu }^{\mathrm{Phys}%
}(p,q)$ and its Borel transformed form contain two structures $\sim p_{\mu }$
and $\sim q_{\mu }$. In investigations we employ the invariant amplitude
that correspond to the structure $\sim p_{\mu }$
\begin{eqnarray}
&&\Pi ^{\mathrm{Phys}}\left( M_{1}^{2},\ M_{2}^{2}\right) =g_{a_{1}f_{0}\pi
}m_{f_{0}}m_{a_{1}}f_{f_{0}}f_{a_{1}}  \notag \\
&&\times \frac{1}{2}e^{-m_{f_{0}}^{2}/M_{1}^{2}-m_{a_{1}}^{2}/M_{2}^{2}}%
\left( -1+\frac{m_{f_{0}}^{2}}{m_{a_{1}}^{2}}\right) .
\end{eqnarray}%
In order to derive the sum rule we need to calculate its second component,
which means that the correlation function $\Pi _{\mu }(p,q)$ has to be found
in terms of quark propagators and distribution amplitudes of the pion. After
substituting the currents into Eq.\ (\ref{eq:CorrF3}) and contracting quark
fields we get
\begin{widetext}

\begin{eqnarray}
&&\Pi _{\mu }^{\mathrm{OPE}}(p,q)=i\int d^{4}x\epsilon \widetilde{\epsilon }%
\epsilon ^{\prime }\widetilde{\epsilon }^{\prime }e^{ipx}\left\{ \mathrm{Tr}%
\left[ \gamma _{\mu }\widetilde{S}_{u}^{a^{\prime }a}(x){}\gamma _{5}%
\widetilde{S}_{s}^{b^{\prime }b}(x){}\right] \left[ \gamma _{5}\widetilde{S}%
_{s}^{ee^{\prime }}(-x)\gamma _{5}{}\right] _{\alpha \beta }\langle \pi (q)|%
\overline{u}_{\alpha }^{c^{\prime }}(x)u_{\beta }^{c}(0)|0\rangle \right.
\notag \\
&&-\mathrm{Tr}\left[ \gamma _{5}\widetilde{S}_{s}^{ee^{\prime }}(-x){}\gamma
_{5}\widetilde{S}_{u}^{cc^{\prime }}(-x){}\right] \left[ \gamma _{\mu }%
\widetilde{S}_{s}^{b^{\prime }b}(x)\gamma _{5}{}\right] _{\alpha \beta
}\langle \pi (q)|\overline{u}_{\alpha }^{a}(x)u_{\beta }^{a^{\prime
}}(0)|0\rangle +\mathrm{Tr}\left[ \gamma _{5}\widetilde{S}_{u}^{a^{\prime
}a}(x){}\gamma _{5}\widetilde{S}_{s}^{b^{\prime }b}(x)\right]  \notag \\
&&\times \left[ \gamma _{5}\widetilde{S}_{s}^{ee^{\prime }}(-x)\gamma _{\mu }%
\right] _{\alpha \beta }\langle \pi (q)|\overline{u}_{\alpha }^{c^{\prime
}}(x)u_{\beta }^{c}(0)|0\rangle +\mathrm{Tr}\left[ \gamma _{5}\widetilde{S}%
_{s}^{ee^{\prime }}(-x)\gamma _{\mu }\widetilde{S}_{u}^{cc^{\prime }}(-x){}%
\right] \left[ \gamma _{5}\widetilde{S}_{s}^{b^{\prime }b}(x)\gamma _{5}{}%
\right] _{\alpha \beta }\langle \pi (q)|\overline{u}_{\alpha
}^{a}(x)u_{\beta }^{a^{\prime }}(0)|0\rangle ,  \notag \\
&&{}  \label{eq:CorrF4}
\end{eqnarray}

\end{widetext}
where for brevity we use $\epsilon \widetilde{\epsilon }\epsilon ^{\prime }%
\widetilde{\epsilon }^{\prime }=\epsilon ^{dab}\epsilon ^{dce}\epsilon
^{d^{\prime }a^{\prime }b^{\prime }}\epsilon ^{d^{\prime }c^{\prime
}e^{\prime }}$.

In this expression $S_{u}(x)$ and $S_{s}(x)$ are the light $u$ and $s$
quarks light-cone propagators (see, Appendix). Let us emphasize that Eq.\ (%
\ref{eq:CorrF4}) is a whole expression for the correlation function, which
encompasses terms appearing due to both $u$ and $d$ components of the
interpolating currents $J_{\mu }(x) $ and $J^{f}(x)$. Because we work in the
chiral limit this form of $\Pi _{\mu }^{\mathrm{OPE}}(p,q)$ is convenient
for further analysis and calculations.

The function $\Pi _{\mu }^{\mathrm{OPE}}(p,q)$ apart from propagators
contains also non-local quark operators sandwiched between the vacuum and
pion states. These operators can be expanded over the full set of Dirac
matrices $\Gamma ^{j}$
\begin{equation}
\overline{u}_{\alpha }^{a}(x)u_{\beta }^{b}(0)\rightarrow \frac{1}{4}\Gamma
_{\beta \alpha }^{j}\left[ \overline{u}^{a}(x)\Gamma ^{j}u^{b}(0)\right] ,
\label{eq:MatEx}
\end{equation}%
where $\Gamma ^{j}$
\begin{equation}
\Gamma ^{j}=\mathbf{1,\ }\gamma _{5},\ \gamma _{\mu },\ i\gamma _{5}\gamma
_{\mu },\ \sigma _{\mu \nu }/\sqrt{2}.
\end{equation}%
Using the projector onto a color -singlet state $\delta ^{ab}/3$ one finds

\begin{equation}
\overline{u}_{\alpha }^{a}(x)u_{\beta }^{b}(0)\rightarrow \frac{\delta ^{ab}%
}{12}\Gamma _{\beta \alpha }^{j}\left[ \overline{u}(x)\Gamma ^{j}u(0)\right]
.
\end{equation}%
The matrix elements of operators $\overline{u}(x)\Gamma ^{j}u(0)$ can be
expanded over $x^{2}$ and expressed by means of the pion's two-particle DAs
of different twist \cite{Braun:1989iv,Ball:1998je,Ball:2006wn}. For example,
in the case of $\Gamma =\ i\gamma _{\mu }\gamma _{5}$ and $\gamma _{5}$ one
obtains%
\begin{eqnarray}
&&\sqrt{2}\langle \pi ^{0}(q)|\overline{u}(x)i\gamma _{\mu }\gamma
_{5}u(0)|0\rangle  \notag \\
&=&f_{\pi }q_{\mu }\int_{0}^{1}due^{i\overline{u}qx}\left[ \phi _{\pi }(u)+%
\frac{m_{\pi }^{2}x^{2}}{16}\mathbb{A}_{4}(u)\right]  \notag \\
&&+\frac{f_{\pi }m_{\pi }^{2}}{2}\frac{x_{\mu }}{qx}\int_{0}^{1}due^{i%
\overline{u}qx}\mathbb{B}_{4}(u),  \label{eq:LTDA}
\end{eqnarray}%
and%
\begin{eqnarray}
&&\sqrt{2}\langle \pi ^{0}(q)|\overline{u}(x)i\gamma _{5}u(0)|0\rangle =%
\frac{f_{\pi }m_{\pi }^{2}}{m_{u}+m_{d}}  \notag \\
&&\times \int_{0}^{1}due^{iuqx}\phi _{3;\pi }^{p}(u).  \label{eq:TW3}
\end{eqnarray}%
In Eq.\ (\ref{eq:LTDA}) $\phi _{\pi }(u)$ is the leading twist (twist-2)
distribution amplitude of the pion, whereas $\mathbb{A}_{4}(u)$ and $\mathbb{%
B}_{4}(u)$ are higher-twist functions that can be expressed using the pion
two-particle twist-4 DAs. One of two-particle twist-3 distributions $\phi
_{3;\pi }^{p}(u)$ determines the matrix element given by Eq.\ (\ref{eq:TW3}%
). \ Another two-particle twist-3 DA $\phi _{3;\pi }^{\sigma }(u)$
corresponds to matix element with $\sigma _{\mu \nu }$ insertion. The
non-local operators that appear due to a gluon field strength tensor $G_{\mu
\nu }(ux)$ included into $\overline{u}(x)\Gamma ^{j}u(0)$ generate the
pion's three-particle distributions. Their definitions and further details
were presented in Ref.\ \cite{Braun:1989iv,Ball:1998je,Ball:2006wn}.

The explicit expression of correlation function in terms of numerous DAs of
the pion is rather cumbersome, therefore we do not provide it here. The $\Pi
_{\mu }^{\mathrm{OPE}}(p,q)$ contains two Lorentz structures $\sim p_{\mu }$
and $\sim q_{\mu }$. We employ the invariant amplitude $\sim p_{\mu }$ to
match it with corresponding function from $\Pi _{\mu }^{\mathrm{Phys}}(p,q)$%
. The Borel transform of the invariant amplitude under discussion, which we
denote in what follows as $\Pi ^{\mathrm{OPE}}\left( M_{1}^{2},\
M_{2}^{2}\right) $, can be calculated along a line explained in Ref.\ \cite%
{Belyaev:1994zk}. At the next phase one must subtract contribution of higher
resonances and continuum states. This procedure becomes easier when two
Borel parameters are equal to each other $M_{1}^{2}=M_{2}^{2}$. In our case
we suggest that a choice $M_{1}^{2}=M_{2}^{2}$ does not generate large
uncertainties in sum rules and introduce $M^{2}$ through
\begin{equation}
\frac{1}{M^{2}}=\frac{1}{M_{1}^{2}}+\frac{1}{M_{2}^{2}},  \label{eq:M1M2}
\end{equation}%
which considerably simplifies studies. Continuum subtraction is fulfilled in
accordance with methods described in Ref.\ \cite{Belyaev:1994zk}. Some
formulas used in this process can be found in Appendix B of Ref.\ \cite%
{Agaev:2016srl}.

Then sum rule for the strong coupling $g_{a_{1}f_{0}\pi }$ reads%
\begin{equation}
g_{a_{1}f_{0}\pi }=\frac{2m_{a_{1}}^{2}}{m_{f_{0}}^{2}-m_{a_{1}}^{2}}\frac{%
e^{(m_{f_{0}}^{2}+m_{a_{1}}^{2})/2M^{2}}}{%
m_{f_{0}}m_{a_{1}}f_{f_{0}}f_{a_{1}}}\Pi ^{\mathrm{OPE}}\left( M^{2},\
s_{0}\right) .  \label{eq:StrCoup1}
\end{equation}%
The partial width of decay $a_{1}(1420)\rightarrow f_{0}(980)\pi ^{0}$ is
given by the formula%
\begin{equation}
\Gamma (a_{1}\rightarrow f_{0}\pi ^{0})=g_{a_{1}f_{0}\pi }^{2}\frac{|%
\overrightarrow{p}|^{3}}{24\pi m_{a_{1}}^{2}},
\end{equation}%
where
\begin{eqnarray}
|\overrightarrow{p}| &=&\frac{1}{2m_{a_{1}}}\left(
m_{a_{1}}^{4}+m_{f_{0}}^{4}+m_{\pi }^{4}-2m_{f_{0}}^{2}m_{a_{1}}^{2}\right.
\notag \\
&&\left. -2m_{\pi }^{2}m_{a_{1}}^{2}-2m_{f_{0}}^{2}m_{\pi }^{2}\right)
^{1/2}.
\end{eqnarray}

Important nonperturbative information in $\Pi ^{\mathrm{OPE}}\left( M^{2},\
s_{0}\right) $ is encoded by the pion's distribution amplitudes. A
substantial part of $\Pi ^{\mathrm{OPE}}\left( M^{2},\ s_{0}\right) $ forms
due to two-particle DAs including $\phi _{\pi }(u)$ one, at the middle point
$u_{0}=1/2$. The leading twist DA $\phi _{\pi }(u)$ through equations of
motion affects other DAs of the pion. As a result, it contributes to $\Pi ^{%
\mathrm{OPE}}\left( M^{2},\ s_{0}\right) $ not only directly, but also
through higher-twist DAs of the pion, and deserves a detailed consideration.

The DA $\phi _{\pi }(u)$ has a following expansion over the Gegenbauer
polynomials $C_{2n}^{3/2}(\varsigma )$
\begin{equation}
\phi _{\pi }(u,\mu ^{2})=6u\overline{u}\left[ 1+\sum_{n=1,2,3\ldots
}a_{2n}(\mu ^{2})C_{2n}^{3/2}(u-\overline{u})\right] ,  \label{eq:PionDALT}
\end{equation}%
where $\overline{u}=1-u$. In general, due to coefficients $a_{2n}(\mu ^{2})$
it depends on a scale $\mu $, as well. Values of the Gegenbauer moments $%
a_{2n}(\mu ^{2})$ at some normalization point $\mu =\mu _{0}$ have to be
either extracted from phenomenological analysis or evaluated by employing,
for example, lattice simulations.

In the present work we use two models for $\phi _{\pi }(u,\mu ^{2}=1\
\mathrm{GeV}^{2}\ )$ . First of them was extracted in Ref. \cite%
{Agaev:2010aq,Agaev:2012tm} from LCSR analysis of the electromagnetic
transition form factor of the pion. This DA is determined by the Gegenbauer
moments
\begin{equation}
a_{2}=0.1,\ a_{4}=0.1,\ a_{6}=0.1,\ a_{8}=0.034,  \label{eq:OurGM}
\end{equation}
and at the middle point equals to $\phi _{\pi }(1/2)\simeq 1.354$. This
value is not far from $\phi _{\mathrm{asy}}(1/2)=3/2$, where $\phi _{\mathrm{%
asy}}(u)=6u\overline{u}$ is the asymptotic DA. We also employ the second
model for $\phi _{\pi }(u)$ obtained in Ref.\ \cite{Braun:2015axa} from
lattice simulations. It contains only one non-asymptotic term
\begin{equation}
\phi _{\pi }(u,\mu ^{2})=6u\overline{u}\left[ 1+a_{2}(\mu ^{2})C_{2}^{3/2}(u-%
\overline{u})\right] ,  \label{eq:PionDALT1}
\end{equation}%
the second Gegenbauer moment $a_{2}(\mu ^{2})$ of which at $\mu =2\ \mathrm{%
GeV}$ was estimated as $a_{2}=0.1364\pm 0.021$. It evolves to
\begin{equation}
a_{2}(1\ \mathrm{GeV}^{2})=0.1836\pm 0.0283,  \label{eq:BRLat}
\end{equation}%
at the scale $\mu =1\ \mathrm{GeV}$.

The sum rule (\ref{eq:StrCoup1}) depends on $a_{1}$ and $f_{0}$ states'
masses and couplings. The parameters $m_{a_{1}}$ and $f_{a_{1}}$ have been
found in the previous section. For $m_{f_{0}}$ we use its value from Ref.\
\cite{Patrignani:2016xqp}
\begin{equation}
m_{f_{0}}=(990\pm 20)\ \mathrm{MeV},
\end{equation}%
whereas the current coupling of the $f_{0}(980)$ meson is borrowed from
Ref.\ \cite{Brito:2004tv}%
\begin{equation}
f_{f_{0}}=(1.51\pm 0.14)\cdot 10^{-3}\ \mathrm{GeV}^{4}.  \label{eq:f0coupl}
\end{equation}%
In Ref.\ \cite{Brito:2004tv} $f_{f_{0}}$ was obtained from QCD sum rules by
employing the interpolating current (\ref{eq:Curr3}), and therefore is
appropriate for our purposes. In Eq.\ (\ref{eq:f0coupl}) we take into
account a difference between definitions of $f_{f_{0}}$ accepted in Ref.\
\cite{Brito:2004tv} and used in the present work.

In computations the Borel and continuum threshold parameters are varied
within the working windows
\begin{equation}
M^{2}\in \lbrack 1.5,\ 2.0]\ \mathrm{GeV}^{2},\ s_{0}\in \lbrack 2.4,\ 3.1]\
\mathrm{GeV}^{2}.
\end{equation}%
In these regions the sum rule complies standard constraints and can be used
to evaluate the strong coupling $g_{a_{1}f_{0}\pi }$ . In Fig.\ \ref%
{fig:StrongC} $g_{a_{1}f_{0}\pi }$ is plotted as a function of the Borel and
continuum threshold parameters. It is seen, that $g_{a_{1}f_{0}\pi }$
demonstrates a nice stability upon varying $M^{2}$, but is sensitive to a
choice of $s_{0}$. Nevertheless, theoretical errors remain within limits
typical for such kind of calculations and do not exceed $30\%$.

For the strong coupling $g_{a_{1}f_{0}\pi }$ and width of the decay $%
a_{1}\rightarrow f_{0}\pi ^{0}$ we find:
\begin{eqnarray}
g_{a_{1}f_{0}\pi } &=&3.41\pm 0.97,  \notag \\
\Gamma (a_{1} &\rightarrow &f_{0}\pi ^{0})=(3.14\pm 0.96)\ \mathrm{MeV,}
\end{eqnarray}%
in the case (\ref{eq:OurGM}), and
\begin{eqnarray}
g_{a_{1}f_{0}\pi } &=&3.38\pm 0.93,  \notag \\
\Gamma (a_{1} &\rightarrow &f_{0}\pi ^{0})=(3.09\pm 0.91)\ \mathrm{MeV,}
\end{eqnarray}%
for the DA defined by Eq.\ (\ref{eq:BRLat}). One can see that an effect on
the final result connected with the choice of the pion leading twist DA is
small.

\begin{widetext}

\begin{figure}[h!]
\begin{center}
\includegraphics[totalheight=6cm,width=8cm]{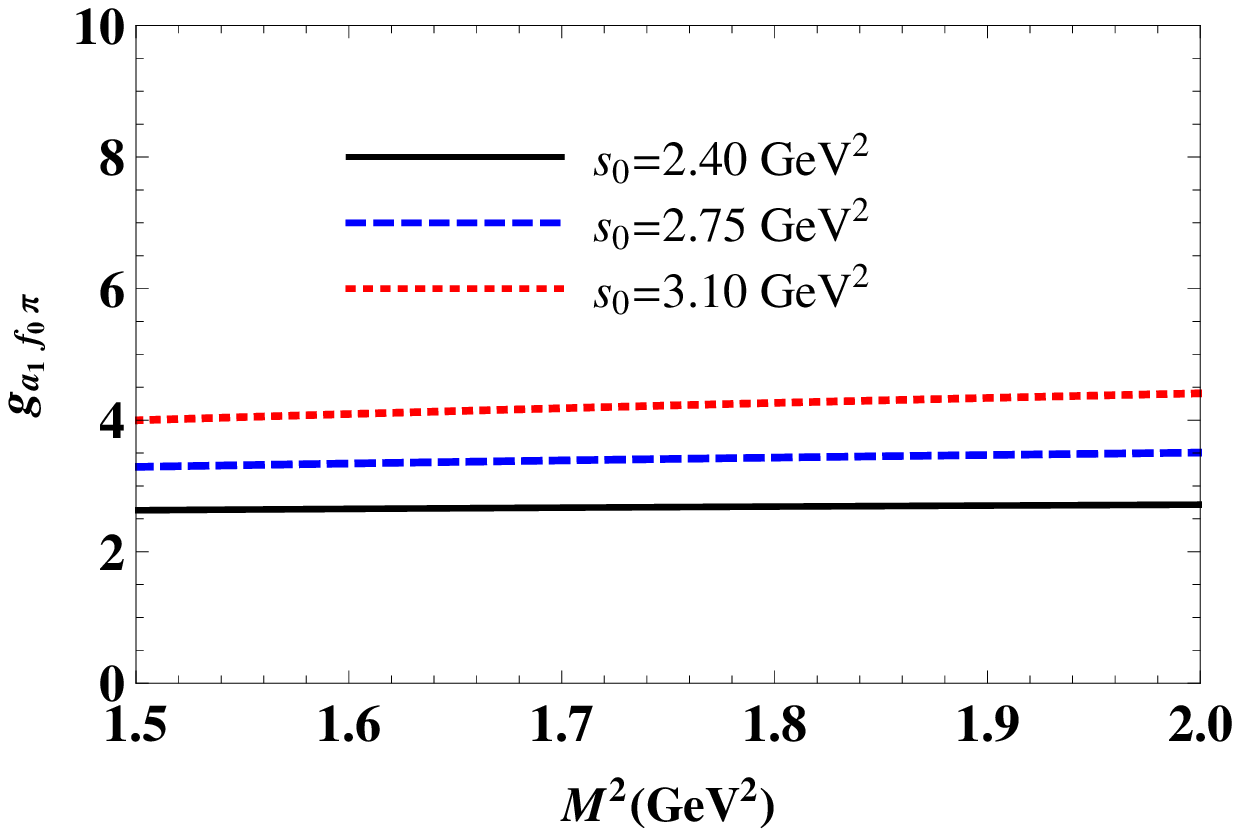}\,\,
\includegraphics[totalheight=6cm,width=8cm]{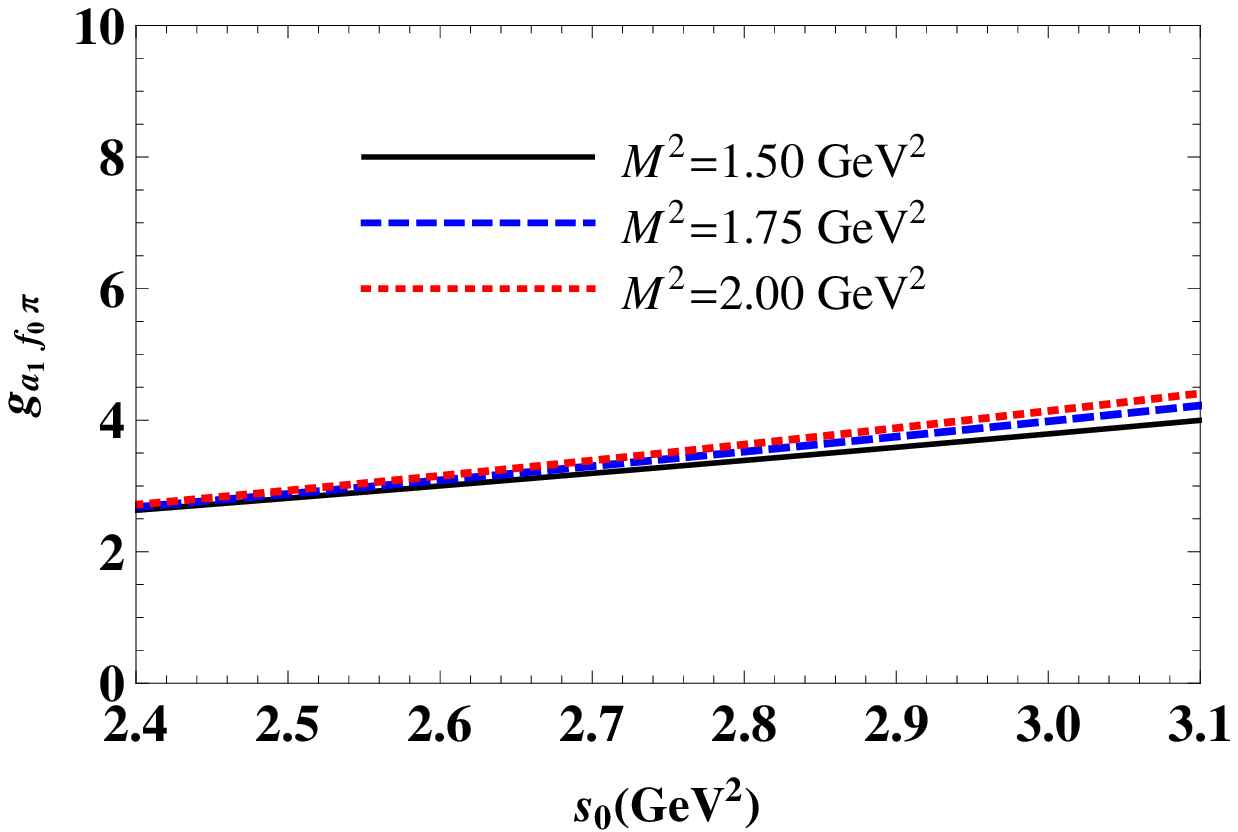}
\end{center}
\caption{ Dependence of the strong coupling $g_{a_1f_0\pi}$  on the Borel parameter $M^2$ at fixed $s_0$
(left panel), and on the continuum threshold $s_0$ at fixed $%
M^2$ (right panel).}
\label{fig:StrongC}
\end{figure}

\end{widetext}


\section{The decay modes $a_1(1420) \to K^{\ast \pm}K^{\mp}$, $K^{\ast 0}%
\bar{K}^{0}$ and $\ \bar{K}^{\ast 0} K^{0}$}

\label{sec:DecK}

In this Section we concentrate on S-wave decays of the $a_{1}(1420)$ state.
To this end we calculate strong couplings $g_{a_{1}K^{\ast }K^{-}}$ , $%
g_{a_{1}K^{\ast }K^{+}}$ and other two ones corresponding to vertices $%
a_{1}K^{\ast 0}\bar{K}^{0}$ and $\ a_{1}\bar{K}^{\ast 0}K^{0}$. All of these
vertices are built of a tetraquark and two conventional mesons. Therefore,
for their exploration, the light-cone sum rule approach has to be
accompanied by the method of the soft meson approximation. This means that
in order to satisfy the four-momentum conservation at these vertices
momentum of a final light meson, for instance, the momentum of $K^{-}$ in $%
a_{1}K^{\ast +}K^{-}$ should be set $q=0$, which leads to important
consequences for a calculational scheme: The distinctive features of the
soft approximation are explained below.

Let us start from decay mode $a_{1}\rightarrow K^{\ast +}K^{-}$ that can be
explored by means of the correlation function
\begin{equation}
\Pi _{\mu \nu }(p,q)=i\int d^{4}xe^{ip\cdot x}\langle K^{-}(q)|\mathcal{T}%
\{J_{\mu }^{K^{\ast +}}(x)J_{\nu }^{\dagger }(0)\}|0\rangle ,
\end{equation}%
where $J_{\mu }^{K^{\ast +}}(x)$ is the correlation function of the $K^{\ast
+}$ meson and has following form
\begin{equation}
J_{\mu }^{K^{\ast +}}(x)=\overline{s}(x)\gamma _{\mu }u(x).
\end{equation}%
Following standard prescriptions we write $\Pi _{\mu \nu }(p,q)$ in terms of
physical parameters of the particles $a_{1},\ K^{\ast +}$ and $K^{-}$
\begin{eqnarray}
&&\Pi _{\mu \nu }^{\mathrm{Phys}}(p,q)=\frac{\langle 0|J_{\mu }^{K^{\ast
+}}|K^{\ast +}(p)\rangle }{p^{2}-m_{K^{\ast }}^{2}}\langle K^{\ast +}\left(
p\right) K^{-}(q)|a_{1}(p^{\prime })\rangle  \notag \\
&&\times \frac{\langle a_{1}(p^{\prime })|J_{\nu }^{\dagger }|0\rangle }{%
p^{\prime 2}-m_{a_{1}}^{2}}+...,  \label{eq:PhysDec2}
\end{eqnarray}%
where $p^{\prime }$ and $p$, $q$ are momenta of the initial and final
particles. In Eq.\ (\ref{eq:PhysDec2}) by dots we show contributions of
excited resonances and continuum states. Further simplification of $\Pi
_{\mu \nu }^{\mathrm{Phys}}(p,q)$ is achieved by utilizing the matrix
elements:
\begin{eqnarray}
&&\langle 0|J_{\mu }^{K^{\ast +}}|K^{\ast +}(p)\rangle =f_{K^{\ast
}}m_{K^{\ast }}\varepsilon _{\mu },  \notag \\
&&\langle K^{\ast +}\left( p\right) K^{-}(q)|a_{1}(p^{\prime })\rangle
=g_{a_{1}K^{\ast }K^{-}}\left[ (p\cdot p^{\prime })(\varepsilon ^{\ast
}\cdot \varepsilon ^{\prime })\right.  \notag \\
&&\left. -(p\cdot \varepsilon ^{\prime })(p^{\prime }\cdot \varepsilon
^{\ast })\right] .
\end{eqnarray}%
First of them, i.e. $\langle 0|J_{\mu }^{K^{\ast +}}|K^{\ast +}(p)\rangle $
is expressed in terms of $K^{\ast +}$ meson's mass $m_{K^{\ast }}$ , decay
constant $f_{K^{\ast }}$ and polarization vector $\varepsilon _{\mu }$. The
second matrix element is written down using the strong coupling $%
g_{a_{1}K^{\ast }K^{-}}$ that should be evaluated from sum rules. In the
soft limit $q\rightarrow 0$ we get $p^{\prime }=p$, as a result have to
perform one-variable Borel transformation, which leads to
\begin{eqnarray}
&&\mathcal{B}\Pi _{\mu \nu }^{\mathrm{Phys}}(p)=g_{a_{1}K^{\ast
}K^{-}}m_{K^{\ast }}m_{a_{1}}f_{K^{\ast }}f_{a_{1}}\frac{e^{-m^{2}/M^{2}}}{%
M^{2}}  \notag \\
&&\times \left( m^{2}g_{\mu \nu }-p_{\nu }p_{\mu }^{\prime }\right) +\ldots ,
\label{eq:BorelPhys}
\end{eqnarray}%
where
\begin{equation}
m^{2}=\frac{m_{K^{\ast }}^{2}+m_{a_{1}}^{2}}{2}.
\end{equation}%
We keep in Eq.\ (\ref{eq:BorelPhys}) $p_{\nu }\neq p_{\mu }^{\prime }$ to
demonstrate explicitly its Lorentz structures. Contributions of higher
resonances and continuum states are denoted in Eq.\ (\ref{eq:BorelPhys}) by
dots: Some of them are not suppressed even after the Borel transformation.
Unsuppressed terms correspond to vertices containing excited states of
involved particles. This is an advantage when one is interested in decays of
excited tetraquarks to conventional ground-state mesons or their decays to
final excited mesons. But it emerges as a problem if one considers only
ground-state vertices. In other words, in the soft limit the
phenomenological side of sum rules takes a complicated form which is one of
aforementioned properties of this approximation \cite{Belyaev:1994zk}.

In the soft limit the correlation function $\Pi _{\mu \nu }^{\mathrm{OPE}%
}(p) $ is given by the formula
\begin{eqnarray}
&&\Pi _{\mu \nu }^{\mathrm{OPE}}(p)=i\int d^{4}xe^{ipx}\frac{\epsilon
\widetilde{\epsilon }}{\sqrt{2}}\left\{ \left[ \gamma _{5}\widetilde{S}%
_{u}^{ic}(x){}\gamma _{\mu }\widetilde{S}_{s}^{bi}(-x){}\gamma _{\nu }\right]
\right.  \notag \\
&&\left. +\left[ \gamma _{\nu }\widetilde{S}_{u}^{ic}(x)\gamma _{\mu }%
\widetilde{S}_{s}^{bi}(-x)\gamma _{5}\right] \right\} _{\alpha \beta
}\langle K^{-}(q)|\overline{s}_{\alpha }^{e}(0)u_{\beta }^{a}(0)|0\rangle ,
\notag \\
&&{}  \label{eq:OPE1}
\end{eqnarray}%
where $\epsilon \widetilde{\epsilon }=\epsilon ^{dab}\epsilon ^{dec}$ . As
is seen, \ $\Pi _{\mu \nu }^{\mathrm{OPE}}(p)$ depends only on local matrix
elements of $K^{-}$ meson.

Contribution to the correlation function comes from the matrix element
\begin{equation}
\langle 0|\overline{u}(0)i\gamma _{5}s(0)|K\rangle =f_{K}\mu _{K},
\end{equation}%
where $\mu _{K}=m_{K}^{2}/(m_{s}+m_{u})$. The function $\Pi _{\mu \nu }^{%
\mathrm{OPE}}(p)$ contains the same Lorentz structures as its
phenomenological counterpart (\ref{eq:BorelPhys}). We work with invariant
amplitudes $\sim g_{\mu \nu }$, and our result for the Borel transform of
the relevant invariant function $\Pi ^{\mathrm{OPE}}(p^{2})$ reads%
\begin{eqnarray}
&&\Pi ^{\mathrm{OPE}}(M^{2})=\int_{4m_{s}^{2}}^{\infty }ds\rho ^{\mathrm{%
pert.}}(s)e^{-s/M^{2}}+\frac{f_{K}\mu _{K}}{\sqrt{2}}  \notag \\
&&\times \left[ \frac{m_{s}}{6}\left( 2\langle \overline{u}u\rangle -\langle
\overline{s}s\rangle \right) \right. +\frac{1}{72}\langle \frac{\alpha
_{s}G^{2}}{\pi }\rangle  \notag \\
&&\left. +\frac{m_{s}}{36M^{2}}\langle \overline{s}g_{s}\sigma Gs\rangle -%
\frac{g_{s}^{2}}{243M^{2}}\left( \langle \overline{s}s\rangle ^{2}+\langle
\overline{u}u\rangle ^{2}\right) \right] ,  \label{eq:Borel2}
\end{eqnarray}%
where
\begin{equation*}
\rho ^{\mathrm{pert.}}(s)=\frac{f_{K}\mu _{K}}{12\sqrt{2}\pi ^{2}}s.
\end{equation*}%
In Eq.\ (\ref{eq:Borel2}) the spectral density $\rho ^{\mathrm{pert.}}(s)$
is found from the imaginary part of the relevant piece in the correlation
function, whereas the Borel transform of other terms are calculated directly
from $\Pi ^{\mathrm{OPE}}(p^{2})$. The function $\Pi ^{\mathrm{OPE}}(M^{2})$
contains terms up to dimension six and has a rather compact form. It is
evident that the soft approximation considerably simplifies the correlation
function $\Pi ^{\mathrm{OPE}}(p^{2})$ , which is another peculiarity of the
method.

The sum rule for the strong coupling $g_{a_{1}K^{\ast }K^{-}}$ should be
obtained from the equality
\begin{equation}
g_{a_{1}K^{\ast }K^{-}}m_{K^{\ast }}m_{a_{1}}f_{K^{\ast }}f_{a_{1}}m^{2}%
\frac{e^{-m^{2}/M^{2}}}{M^{2}}+...=\Pi ^{\mathrm{OPE}}(M^{2}).
\label{eq:SRraw}
\end{equation}%
But before performing the standard continuum subtraction one needs to remove
unsuppressed terms from the left-hand side of this equality. This task is
solved by acting on both sides of Eq.\ (\ref{eq:SRraw}) by the operator \cite%
{Belyaev:1994zk,Ioffe:1983ju}
\begin{equation*}
\mathcal{P}(M^{2},m^{2})=\left( 1-M^{2}\frac{d}{dM^{2}}\right)
M^{2}e^{m^{2}/M^{2}},
\end{equation*}%
that singles out the ground-state term. Remaining contributions can be
subtracted afterwards in a usual manner, which requires the replacing (\ref%
{eq:SubrtF}) in the first term of \ $\Pi ^{\mathrm{OPE}}(M^{2})$ while
leaving components $\sim (M^{2})^{0}$ and $\sim 1/M^{2}$ in their original
forms \cite{Belyaev:1994zk}.

The width of the decay $a_{1}\rightarrow K^{\ast +}K^{-}$ is given by the
following expression%
\begin{equation*}
\Gamma (a_{1}\rightarrow K^{\ast +}K^{-})=\frac{g_{a_{1}K^{\ast
}K^{-}}^{2}m_{K^{\ast }}^{2}}{24\pi }|\overrightarrow{p}|\left( 3+\frac{2|%
\overrightarrow{p}|^{2}}{m_{K^{\ast }}^{2}}\right) ,
\end{equation*}%
where now
\begin{eqnarray}
&&|\overrightarrow{p}|=\frac{1}{2m_{a_{1}}}\left( m_{a_{1}}^{4}+m_{K^{\ast
}}^{4}+m_{K}^{4}-2m_{K^{\ast }}^{2}m_{a_{1}}^{2}\right.  \notag \\
&&\left. -2m_{K}^{2}m_{a_{1}}^{2}-2m_{K^{\ast }}^{2}m_{K}^{2}\right) ^{1/2}.
\end{eqnarray}

The obtained for $g_{a_{1}K^{\ast }K^{-}}$ sum rule can be easily adopted
for numerical computations. The working windows for the parameters $M^{2}$
and $s_{0}$ used in the case of the $a_{1}\rightarrow f_{0}(980)\pi $ decay
is suitable for $a_{1}\rightarrow K^{\ast +}K^{-}$ process, as well. The
mass of mesons $K^{\ast +}$ and $K^{-}$ are taken from Ref. \cite%
{Patrignani:2016xqp}
\begin{eqnarray}
m_{K^{\pm }} &=&(493.677\pm 0.016)\ \ \mathrm{MeV},  \notag \\
m_{K^{\ast \pm }} &=&(891.76\pm 0.25)\ \mathrm{MeV}.
\end{eqnarray}%
For their decay constants we use
\begin{eqnarray}
f_{K^{\pm }} &=&(155.72\pm 0.51)\ \mathrm{MeV,}  \notag \\
f_{K^{\ast 0(\pm )}} &=&225\ \mathrm{MeV.}  \label{eq:Kdecays}
\end{eqnarray}%
Results of calculations are presented below:%
\begin{eqnarray}
&&g_{a_{1}K^{\ast }K^{-}}=(2.84\pm 0.79)\ \mathrm{GeV}^{-1},  \notag \\
&&\Gamma (a_{1}\rightarrow K^{\ast +}K^{-})=(37.84\pm 10.97)\ \mathrm{MeV}.
\label{eq:DW2}
\end{eqnarray}%
The strong coupling $g_{a_{1}K^{\ast }K^{+}}$ and width of the decay $\Gamma
(a_{1}\rightarrow K^{\ast -}K^{+})$ are also given by Eq.\ (\ref{eq:DW2}).

The analysis of the next two partial decay channels of the $a_{1}(1420)$
state $a_{1}\rightarrow K^{\ast 0}\bar{K}^{0}\left( \bar{K}^{\ast
0}K^{0}\right) $ does not differ from one presented in a detailed form in
this section. Let us only write down masses of the $K^{0}(\bar{K}^{0})$ and $%
K^{\ast 0}(\bar{K}^{\ast 0})$ mesons
\begin{eqnarray}
m_{K^{0}} &=&(497.611\pm 0.013)\ \mathrm{MeV},  \notag \\
m_{K^{\ast 0}} &=&(895.55\pm 0.20)~\mathrm{MeV,}
\end{eqnarray}%
used in numerical computations. The decay constants of these pseudoscalar
and vector mesons are taken equal to ones from Eq.\ (\ref{eq:Kdecays}). We
omit further details and write down final sum rules' predictions for these
channels:
\begin{eqnarray}
&&g_{a_{1}K^{\ast 0}\overline{K}^{0}}=(2.85\pm 0.82)\ \mathrm{GeV}^{-1},
\notag \\
&&\Gamma (a_{1}\rightarrow K^{\ast 0}\bar{K}^{0})=(33.35\pm 9.76)\ \mathrm{%
MeV}.
\end{eqnarray}%
The obtained in the sections \ref{sec:Decf0} and \ref{sec:DecK} results for
partial decays of the $a_{1}(1420)$ state enable us to calculate its full
width: For $\Gamma $ we get
\begin{equation}
\Gamma =(145.52\pm 20.79)\text{ }\mathrm{MeV,}
\end{equation}%
which within theoretical errors of sum rule computations is compatible with
the data provided by the COMPASS Collaboration.


\section{Concluding notes}

\label{sec:Notes}

The $a_{1}(1420)$ meson recently discovered by the COMPASS Collaboration
took its place in an already overpopulated range of the light $a_{1}$
axial-vector mesons with $J^{PC}=1^{++}$ worsening a situation with their
interpretation. The standard model of the mesons as bound states of a quark
and a antiquark meets with difficulties to find a proper place for all of
them. Thus, the $a_{1}(1420)$ meson can not be interpreted as the radial
excitation of $a_{1}(1260)$, because the mass difference between them is
small to accept this assumption. Explanations of $a_{1}(1420)$ as an
dynamical effect observed in $a_{1}(1260)\rightarrow f_{0}(980)\pi $ decay
are presented in the literature by number of works.

Alternative interpretations of the $a_{1}(1420)$ meson as a four-quark
exotic state are among models in use, as well. In the framework of this
approach it was considered as a pure diquark-antidiquark state or some
admixture of a diquark-antidiquark and $q\overline{q}$ component. The
molecular organization for $a_{1}(1420)$ was also employed in the literature.

In the present work we have treated the $a_{1}(1420)$ meson as a pure
diquark-antidiquark state, and calculated its spectroscopic parameters, and
widths of five decay modes. Our result for the mass of $a_{1}(1420)$
evaluated using QCD two-point sum rule method and given by $%
m_{a_{1}}=1416_{-79}^{+81}\ \ \mathrm{MeV}$ is in excellent agreement with
the experimental result. It is also  in accord within small computational errors
with the prediction obtained in Ref.\ \cite{Chen:2015fwa}. The full width of
the $a_{1}(1420)$ meson has been calculated on basis of its five decay
channels and led to $\Gamma =(145.52\pm 20.79)$ $\mathrm{MeV}$. By taking
into account theoretical uncertainties of our calculations and experimental
errors it agrees with measurements by the COMPASS Collaboration $\Gamma
=153_{-23}^{+8}\ \mathrm{MeV}$.

Present studies have confirmed that the $a_{1}(1420)$ meson can be
considered as a serious candidate to an exotic state, and a
diquark-antidiquark model of its structure deserves detailed investigations.


\section*{ACKNOWLEDGEMENTS}

K.~A.~ thanks TUBITAK for the partial financial support provided under Grant
No. 115F183.

\appendix*

\section{ The light quark propagators and $\Pi _{\mathrm{V}}(M^{2},\ s_{0})$}

\renewcommand{\theequation}{\Alph{section}.\arabic{equation}} \label{sec:App}
The light quark propagator is necessary to find QCD side of the correlation
functions in the mass and strong couplings' calculations. In the present
work for the light $q$-quark propagator $S_{q}^{ab}(x)$ we use the following
formula
\begin{widetext}
\begin{eqnarray}
&&S_{q}^{ab}(x)=i\delta _{ab}\frac{\slashed x}{2\pi ^{2}x^{4}}-\delta _{ab}%
\frac{m_{q}}{4\pi ^{2}x^{2}}-\delta _{ab}\frac{\langle \overline{q}q\rangle
}{12}+i\delta _{ab}\frac{\slashed xm_{q}\langle \overline{q}q\rangle }{48}%
-\delta _{ab}\frac{x^{2}}{192}\langle \overline{q}g_{s}\sigma Gq\rangle
+i\delta _{ab}\frac{x^{2}\slashed xm_{q}}{1152}\langle \overline{q}%
g_{s}\sigma Gq\rangle   \notag \\
&&-i\frac{g_{s}G_{ab}^{\alpha \beta }}{32\pi ^{2}x^{2}}\left[ \slashed x{%
\sigma _{\alpha \beta }+\sigma _{\alpha \beta }}\slashed x\right] -i\delta
_{ab}\frac{x^{2}\slashed xg_{s}^{2}\langle \overline{q}q\rangle ^{2}}{7776}%
-\delta _{ab}\frac{x^{4}\langle \overline{q}q\rangle \langle
g_{s}^{2}G^{2}\rangle }{27648}+\ldots   \label{eq:qprop}
\end{eqnarray}%
The propagator (\ref{eq:qprop}) has been used to calculate the correlation
function $\Pi _{\mu \nu }^{\mathrm{OPE}}(p)$. Below we provide the
subtracted Borel transform of $\Pi _{\mathrm{V}}(p^{2})$
\begin{eqnarray}
&&\Pi _{\mathrm{V}}(M^{2},s_{0})=\frac{1}{2}\sum_{q=u,d}\left\{
\int_{4m_{s}^{2}}^{s_{0}}dse^{-s/M^{2}}\left[ \frac{s^{4}}{9\cdot 2^{12}\pi
^{6}}+s^{2}\left( \frac{m_{s}(3\langle \overline{s}s\rangle -7\langle
\overline{q}q\rangle )}{3\cdot 2^{7}\pi ^{4}}+\frac{\langle G^{2}\rangle }{%
9\cdot 2^{9}\pi ^{4}}\right) +s\left( \frac{5m_{s}^{2}\langle G^{2}\rangle }{%
9\cdot 2^{11}\pi ^{4}}+\frac{5\langle \overline{s}s\rangle \langle \overline{%
q}q\rangle }{36\pi ^{2}}\right. \right. \right.   \notag \\
&&\left. -\frac{m_{s}(4\langle \overline{q}g_{s}\sigma Gq\rangle -15\langle
\overline{s}g_{s}\sigma Gs\rangle )}{9\cdot 2^{6}\pi ^{4}}+\frac{%
g_{s}^{2}(\langle \overline{s}s\rangle ^{2}+\langle \overline{q}q\rangle
^{2})}{972\pi ^{4}}\right) +\frac{m_{s}^{2}(\langle \overline{s}s\rangle
^{2}-6\langle \overline{s}s\rangle \langle \overline{q}q\rangle +8\langle
\overline{q}q\rangle ^{2})}{48\pi ^{2}}  \notag \\
&&\left. -\frac{\langle G^{2}\rangle m_{s}(\langle \overline{s}s\rangle
+17\langle \overline{q}q\rangle )}{3\cdot 2^{9}\pi ^{2}}-\frac{\langle
G^{2}\rangle ^{2}}{3\cdot 2^{10}\pi ^{2}}-\frac{\langle \overline{s}%
g_{s}\sigma Gs\rangle \langle \overline{q}q\rangle }{8\pi ^{2}}\right] -%
\frac{m_{s}\langle G^{2}\rangle (2\langle \overline{s}g_{s}\sigma Gs\rangle
-27\langle \overline{q}g_{s}\sigma Gq\rangle )}{27\cdot 2^{9}\pi ^{2}}
\notag \\
&&+\frac{2m_{s}(\langle \overline{s}s\rangle ^{2}\langle \overline{q}%
q\rangle -2\langle \overline{q}q\rangle ^{2}\langle \overline{s}s\rangle )}{9%
}+\frac{m_{s}g_{s}^{2}}{486\pi ^{2}}\left[ \langle \overline{s}s\rangle
^{3}-2\langle \overline{s}s\rangle ^{2}\langle \overline{q}q\rangle
+2\langle \overline{s}s\rangle \langle \overline{q}q\rangle ^{2}-2\langle
\overline{q}q\rangle ^{3}\right] +\frac{\langle \overline{s}g_{s}\sigma
Gs\rangle \langle \overline{q}g_{s}\sigma Gq\rangle }{3\cdot 2^{4}\pi ^{2}}
\notag \\
&&+\frac{\langle G^{2}\rangle \left( \langle \overline{s}s\rangle
^{2}+25\langle \overline{s}s\rangle \langle \overline{q}q\rangle -2\langle
\overline{q}q\rangle ^{2}\right) }{27\cdot 2^{5}}+\frac{g_{s}^{2}\langle
G^{2}\rangle \left( \langle \overline{s}s\rangle ^{2}+\langle \overline{q}%
q\rangle ^{2}\right) }{2^{4}\cdot 3^{6}\pi ^{2}}+\frac{1}{M^{2}}\left[ \frac{%
m_{s}\langle G^{2}\rangle ^{2}(2\langle \overline{s}s\rangle +9\langle
\overline{q}q\rangle )}{81\cdot 2^{10}}+\frac{2m_{s}\langle \overline{s}%
s\rangle ^{2}\langle \overline{q}g_{s}\sigma Gq\rangle }{27}\right.   \notag
\\
&&+\frac{g_{s}^{2}m_{s}(\langle \overline{s}g_{s}\sigma Gs\rangle -\langle
\overline{q}g_{s}\sigma Gq\rangle )}{2^{3}\cdot 3^{6}\pi ^{2}}\left(
2\langle \overline{s}s\rangle ^{2}-\langle \overline{s}s\rangle \langle
\overline{q}q\rangle +3\langle \overline{q}q\rangle ^{2}\right) +\frac{%
\langle G^{2}\rangle \langle \overline{s}g_{s}\sigma Gs\rangle \langle
\overline{q}q\rangle }{432} \notag \\
&&\left. \left. -\frac{g_{s}^{4}}{2^{4}\cdot 3^{7}\pi ^{2}}\left( \langle
\overline{s}s\rangle ^{4}+4\langle \overline{s}s\rangle ^{2}\langle
\overline{q}q\rangle ^{2}+\langle \overline{q}q\rangle ^{4}\right) \right]
\right\} ,  \label{eq:PiV}
\end{eqnarray}%
where we have used the notation
\begin{equation*}
\langle G^{2}\rangle =\langle \frac{\alpha _{s}G^{2}}{\pi }\rangle .
\end{equation*}%
In Section \ref{sec:Decf0} we have used the light-cone propagators of the
light $q=u$ and $s$ quarks. This propagator is determined by the
formula
\begin{eqnarray}
&&S_{q}^{ab}(x)=\frac{i\slashed x}{2\pi ^{2}x^{4}}\delta _{ab}-\frac{m_{q}}{%
4\pi ^{2}x^{2}}\delta _{ab}-\frac{\langle \overline{q}q\rangle }{12}\left(
1-i\frac{m_{q}}{4}\slashed x\right) \delta _{ab}-\frac{x^{2}}{192}%
m_{0}^{2}\langle \overline{q}q\rangle \left( 1-i\frac{m_{q}}{6}\slashed %
x\right) \delta _{ab}  \notag \\
&&-ig_{s}\int_{0}^{1}du\left\{ \frac{\slashed x}{16\pi ^{2}x^{2}}G_{ab}^{\mu
\nu }(ux)\sigma _{\mu \nu }-\frac{iux_{\mu }}{4\pi ^{2}x^{2}}G_{ab}^{\mu \nu
}(ux)\gamma _{\nu }-\frac{im_{q}}{32\pi ^{2}}G_{ab}^{\mu \nu }(ux)\sigma
_{\mu \nu }\left[ \ln \left( \frac{-x^{2}\Lambda ^{2}}{4}\right) +2\gamma
_{E}\right] \right\} ,
\end{eqnarray}%
where $\gamma _{E}\simeq 0.577$ is the Euler constant and $\Lambda $ is the
QCD scale parameter.
\end{widetext}

\end{document}